\documentclass[aps,prl,twocolumn,superscriptaddress,groupedaddress]{revtex4}  
\usepackage{graphicx,tensor, cleveref}  
\usepackage{dcolumn}   
\usepackage{bm}        
\usepackage{amssymb}   
\usepackage{amsmath}
\usepackage{dsfont}
\usepackage{cancel}
\usepackage{braket}
\usepackage{xcolor}

\begin{document}



\title{Detecting qubit entanglement : an alternative to the PPT test}
\author{ Joseph Samuel, Kumar Shivam and  Supurna Sinha}

\address{Raman Research Institute, Bangalore 560080, India.}

\date{\today}

\begin{abstract}
We propose a Partial Lorentz Transformation (PLT) test for detecting entanglement in a two qubit system. One can expand the 
density matrix of a two qubit system in terms of a tensor product of $(\mathbb{I}, \vec{\sigma})$. 
The matrix $A$ of the coefficients that appears in such an expansion can be ``squared" to form a 
$4\times4$ matrix $B$. It can be shown that the eigenvalues $\lambda_0, \lambda_1, \lambda_2, \lambda_3$ of $B$ 
are positive. With the choice of $\lambda_0$ as the dominant eigenvalue,
the separable states satisfy 
$\sqrt{\lambda_1}+\sqrt{\lambda_2}+\sqrt{\lambda_3}\leq \sqrt{\lambda_0}$.
Violation of this inequality is a test of entanglement. 
Thus, this condition is both necessary and sufficient and serves 
as an alternative to the celebrated Positive Partial 
Transpose (PPT) test for entanglement detection. 
We illustrate this test by considering some explicit examples.

\end{abstract}

\maketitle



\section{I. Introduction}
Entanglement is considered a valuable resource in quantum information 
processing. Yet, quantum entanglement is a property surprisingly 
hard to detect. Given an arbitrary mixed state (a density matrix) 
of a bipartite system, is it separable or entangled? The problem in
its general form is considered to be difficult. But in low dimensional
Hilbert spaces, some progress has been made.
The simplest example of quantum entanglement
occurs in the case of two qubits. This example deserves to be thoroughly
understood in order for us to address the harder problems 
of entanglement in higher dimensional quantum systems.

In two qubit systems, the Positive Partial Transpose (PPT) criterion 
\cite{peres,horodeckipla,geometry} 
gives a simple, computable criterion for detecting entanglement. 
The criterion gives a necessary and suficient condition for a 
state to be separable. Here we propose a new test which we call 
the Partial Lorentz Transformation (PLT) test 
for detecting entanglement in a two qubit system.
We claim that the PLT
(like the PPT) is a necessary and 
sufficient condition and serves as an alternative to the PPT criterion. 
Our purpose in this paper is to describe the test in recipe form, so that 
it is readily used. The mathematical theory based on Partial Lorentz Transformation and proofs underlying the test 
are explained elsewhere\cite{lsvd}. 

The paper is organized as follows. In Section II we outline 
the PLT entanglement test for two qubits that we propose here. 
In Section III we illustrate the test by considering a few specific cases. 
Finally we end the paper with some concluding remarks in Section IV.



\section{II. Entanglement Test for Two Qubits}
Let $\rho$ be a density matrix of a two qubit system. 
If $\rho$ can be expressed in the form ($\tau^1$ and $\tau^2$ 
are 1-qubit density matrices)
\begin{equation}\label{sepden}
	\rho = \sum_{i} w_{i}\ \tau_i^1\otimes\tau_i^2 \qquad \qquad w_{i}>0
\end{equation}
we say that $\rho$ is separable. Else $\rho$ is entangled.
We assume $\rho$ is positive ($\rho\ge 0$) and Hermitian ($\rho^{\dagger}=\rho$).
In our treatment, we will not need to normalize $\rho$. 
One can expand the density matrix $\rho$ as
\begin{equation}
        \rho=\frac{1}{4} A^{\mu\nu}\sigma_{\mu}\otimes\sigma_{\nu}
\end{equation}
where $\sigma_\mu = (\mathbb{I},\ \sigma_1,\ \sigma_2,\ \sigma_3)$ are the identity and the Pauli matrices. $A^{\mu\nu}$ can be calculated from
\begin{equation}\label{A}
 A_{\mu\nu}=\textrm{Tr}(\rho\sigma_{\mu}\otimes\sigma_{\nu}).
\end{equation}
Let us consider $B\indices{^{\mu}_{\nu}}=A\indices{^{\mu \alpha}}g_{\alpha\beta}A\indices{^{\sigma \beta}}g_{\sigma\nu}$ where $g=\textrm{diag}(1,-1,-1,-1)$
is the Minkowski metric. Now,
\begin{equation}\label{B}
B\indices{^{\mu}_{\nu}}=A\indices{^{\mu}_{\alpha}}A\indices{_{\nu}^{\alpha}}
\end{equation}
is obviously symmetric $(B\indices{^{\mu\nu}}=B\indices{^{\nu\mu}})$. 
It can be shown\cite{lsvd, supplementary} that the eigenvalues of $B$ \eqref{B} are non-negative and so we can define $\mu_a=\sqrt{\lambda_a}$ to be real where 
$a=0,1,2,3$.

Our claim is that, the necessary and sufficient condition for separability is
	\begin{equation}\label{sec}
T(\mu_a)=\mu_0-(\mu_1+\mu_2+\mu_3)\geq 0.
	\end{equation}
Violation of this inequality signals entanglement.
This is our PLT test for entanglement of two qubits. The PLT test is an alternative to the PPT test which is widely known and used in this field.
\section{III. detection of entanglement: A few Examples}
In this section we consider some specific families of states to illustrate the use of our criterion for detecting entanglement. We first consider the Werner state.

\subsection{Example-I}
The Werner state is a two qubit mixed state given by $\rho^{W} = \frac{1-\alpha}{4}\mathbb{I}+\alpha\ket{\mathcal{S}}\bra{\mathcal{S}}$ where $\ket{\mathcal{S}}=\frac{|\uparrow \downarrow \rangle-|\downarrow \uparrow \rangle }{\sqrt{2}}$ is a spin singlet state and $0\leq \alpha \leq 1$. The PPT test shows that this state is separable for $0 \leq \alpha \leq 1/3$ and  entangled for $1/3<\alpha\leq1$.
Let us first construct $A_{\mu\nu}=\text{Tr}\left[\rho ^W \sigma _{\mu }\otimes \sigma _{\nu }
\right]$. 
\begin{equation}
A\indices{_\mu_\nu}=\left(
\begin{array}{cccc}
	1 & 0 & 0 & 0 \\
	0 & -\alpha  & 0 & 0 \\
	0 & 0 & -\alpha  & 0 \\
	0 & 0 & 0 & -\alpha  \\
\end{array}
\right).
\end{equation}
From $A_{\mu\nu}$ we can construct matrix $B$:

\begin{equation}
	B\indices{^\mu_\nu}=\left(
	\begin{array}{cccc}
	1 & 0 & 0 & 0 \\
	0 & \alpha ^2 & 0 & 0 \\
	0 & 0 & \alpha ^2 & 0 \\
	0 & 0 & 0 & \alpha ^2 \\
	\end{array}
	\right)
\end{equation}
The eigenvalues of $B$ are $\lambda_0=1$ and $\lambda_1,\lambda_2,\lambda_3=\alpha^2$. They are positive as claimed earlier and hence we can take the positive square-root of 
these eigenvalues to obtain the ${\mu}$s. We now apply the test by computing $T(\alpha)= \sqrt{\lambda_0}-\sqrt{\lambda_1}-\sqrt{\lambda_2}-\sqrt{\lambda_3} $. A state is entangled iff $T(\alpha)<0$, which gives

\[1-3\alpha<0 \qquad \Rightarrow \qquad \alpha>\frac{1}{3}
\]
\begin{figure}[h!]
	\begin{center}
		\includegraphics[width=0.45
		 \textwidth]{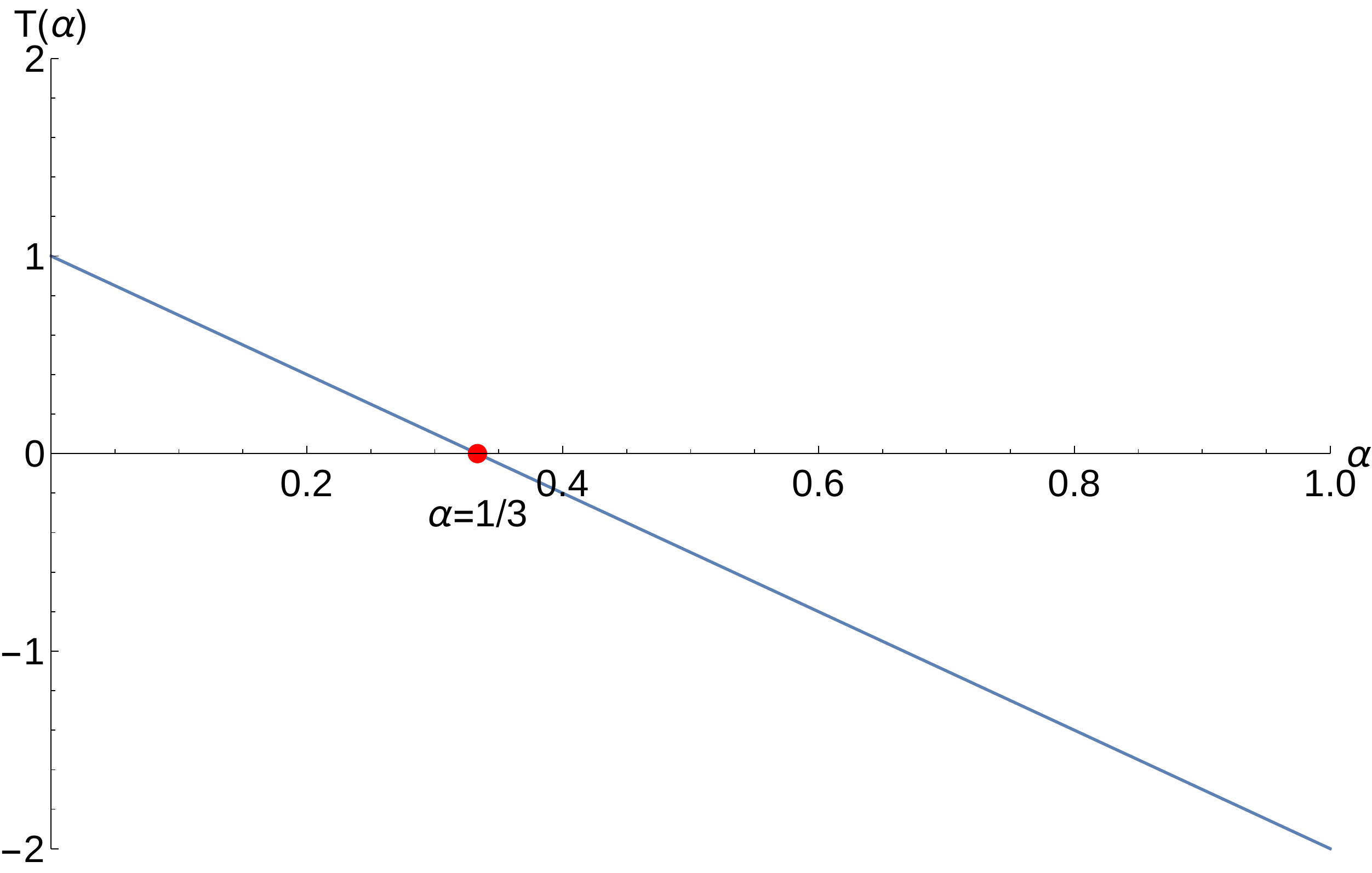}
		\caption{A figure showing the variation of $T(\alpha)$ with $\alpha$. We see that for $\alpha>1/3,$  $T(\alpha)$ is negative and the state is therefore entangled.}
	\end{center}
\end{figure}
Hence, we correctly obtain the condition $1/3<\alpha\leq1$  for $\rho$ to be entangled (See Fig.1).

\subsection{Example-II}
Another interesting example is the state given by\cite{rudolph}
\begin{eqnarray} \label{ccn}
	\rho=&\frac{1}{4}&((r-s+1) \sigma_3\otimes \sigma_3+r(\sigma _3\otimes \mathcal{I})
	+s(\mathcal{I}\otimes \sigma_3)  \nonumber \\
	&+&t(\sigma_1\otimes \sigma_1)-t(\sigma_2\otimes \sigma_2)+\mathcal{I}\otimes \mathcal{I}).
\end{eqnarray}
$\rho$ is a two qubit density matrix for the parameter range
\begin{eqnarray}\label{sc}
s-r&\geq& 0\nonumber \\
|r|&\leq& 1 \nonumber \\ 
|s|&\leq& 1 \nonumber \\ 
t^2&\leq& (1-s)(1+r)\equiv h^2.
\end{eqnarray}

Applying the partial transpose test to $\rho$ we find that the state is entangled\cite{rudolph} for $|t|\neq 0$ and separable for $|t|=0$. Let us apply the PLT test on $\rho$. Following the same recipe as in the previous example we find,
\begin{equation}
A\indices{_\mu_\nu} =	\left(
	\begin{array}{cccc}
	1 & 0 & 0 & s \\
	0 & t & 0 & 0 \\
	0 & 0 & -t & 0 \\
	r & 0 & 0 & r-s+1 \\
	\end{array}
	\right)
\end{equation}

\begin{equation}
	B\indices{^\mu_\nu} = \left(
	\begin{array}{cccc}
	1-s^2 & 0 & 0 & (s-r) (1-s) \\
	0 & t^2 & 0 & 0 \\
	0 & 0 & t^2 & 0 \\
	-(s-r) (1-s) & 0 & 0 & -(2 r-s+1) (s-1) \\
	\end{array}
	\right)
\end{equation}
and the eigenvalues of $B$ are $(h^2,\ h^2,\ t^2,\ t^2)$. The last inequality of the state condition\eqref{sc} implies 
that the dominant eigenvalue of the matrix $B$ is $h^2$. Then the PLT condition for separability requires,
\[|h|\geq |h|+2|t| \implies |t|\leq 0 \implies t=0.
\]
Hence, we find that the state $\rho$ is separable for $t=0$ and entangled otherwise  
which is in agreement with the PPT test.

The computable cross norm (CCN) test proposed in \cite{rudolph} 
doesn't detect entanglement for all states. 
It only works if the reduced density matrices of the individual systems are
maximally disordered ($r=0$ and $s=0$ in Example II)\cite{rudolph}. 
For instance, it does not work for the 
the state $\rho$ given by \eqref{ccn} for the 
parameters $r=1/4$ and $s=1/2$, $t=1/16$. 
However, we notice that the PLT gives $T(t)=-4\sqrt{2/5}\ |t|$ showing that the state 
is not separable for any non-zero value of $t$ (See Fig. 2) in agreement with the PPT test.


\begin{figure}[h!]
	\begin{center}
		\includegraphics[width=0.45
		\textwidth]{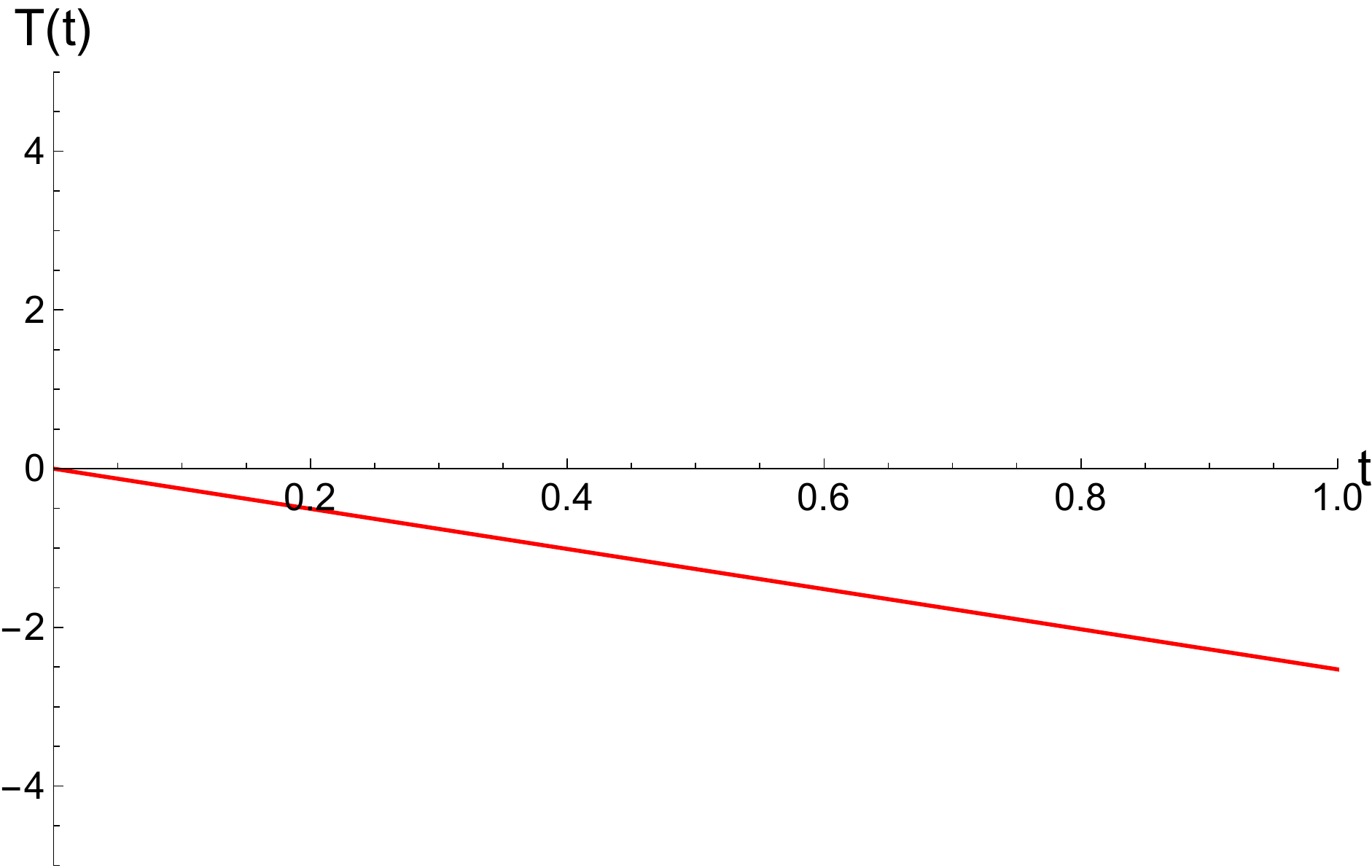}
		\caption{A figure showing the variation of $T(t)$ with $t$. We see that for all $t\in (0,1]$  $T(t)$ is negative, indicating non-separability.}
	\end{center}
\end{figure}

\subsection{Example-III}
Now, we apply this test to a set of states which are incoherent mixtures of the singlet state and the maximally polarized state\cite{peres}.
\begin{equation}
	\rho=x \ket{\mathcal{S}}\bra{\mathcal{S}}+ (1-x)\ket{\uparrow\uparrow}\bra{\uparrow\uparrow}
\end{equation}
where $\ket{\mathcal{S}}$ is the singlet state and $0\leq x \leq 1$. We compute the matrices $A\  \textrm{and}\ B$ and find
\begin{equation}
A\indices{_\mu_\nu}=\left(
\begin{array}{cccc}
1 & 0 & 0 & 1-x \\
0 & -x & 0 & 0 \\
0 & 0 & -x & 0 \\
1-x & 0 & 0 & 1-2 x \\
\end{array}
\right)
\end{equation}
\begin{equation}
B\indices{^\mu_\nu} =	\left(
	\begin{array}{cccc}
	-(x-2) x & 0 & 0 & 2 (x-1) x \\
	0 & x^2 & 0 & 0 \\
	0 & 0 & x^2 & 0 \\
	-2 (x-1) x & 0 & 0 & x (3 x-2) \\
	\end{array}
	\right).
\end{equation}
The eigenvalues of $B$ are $(x^2,\ x^2,\ x^2,\ x^2),$ which are all positive. These states turn out to be entangled since $T(x)=-2x<0$ 
for all values of the parameter $x\in (0,1]$ in agreement with the PPT test.  
\subsection{Example-IV}
Finally, we present the results of a numerical study in support of the claims
made in this paper. A Mathematica program 
(see supplementary material) generates random density 
matrices of a two qubit system. The states are then
tested for entanglement using the PLT test and the PPT test. In {\it all}
cases, we find that the two tests agree. The program also shows by 
numerical evidence that the eigenvalues of $B$ (\ref{B}) are positive. 
Both these claims are supported by a detailed mathematical study
\cite{lsvd}. However, 
the PLT test can 
be profitably used without having to go through the proofs of \cite{lsvd}

\section{IV. Conclusion}
We have presented a necessary and sufficient criterion, to detect two qubit entanglement. This criterion is distinct from the 
celebrated PPT test and thus serves as an alternative method for detection of entanglement. We have explicitly demonstrated 
that this test works for some specific cases. More generally, we numerically generate random density matrices and show 
that the PLT test agrees with the PPT test in all cases \cite{supplementary}.

Expressions similar 
to $T$ have appeared before in \cite{Wootters}, which studies the entanglement of formation for two qubit systems. 
However our work goes beyond this, in proposing an explicit test for detection of two qubit 
entanglement, which serves as an alternative to the PPT test. 

In Ref\cite{rudolph} a separability criterion called the 
computable cross norm (CCN) is proposed. Example-II is taken from \cite{rudolph} and shows that the CCN test 
fails to detect entanglement, while the PPT and PLT succeed. 
Thus the PLT test is a more discriminating test for detection of entanglement compared to the CCN test and just as good as the PPT test. 
It is worth noting that the use of a Lorentzian metric is the crucial ingredient that leads to the sucess of the PLT test. 
We expect that this framework for detection of entanglement can be extended to higher dimensional examples beyond two qubits.  
This test is of relevance to the area of Quantum Information where entanglement is viewed as an important resource. 

\section{Acknowledgements}
We thank Anirudh Reddy for discussions.



\end{document}